\documentclass[twocolumn,aps,prd]{revtex4}
\usepackage{latexsym}
\usepackage{epsfig}

\newcommand{\gsim}{\ \raise.3ex\hbox{$>$\kern-.75em\lower1ex\hbox{$\sim$}} \ }
\newcommand{\lsim}{\ \raise.3ex\hbox{$<$\kern-.75em\lower1ex\hbox{$\sim$}} \ }
\def\be{\begin{equation}}
\def\ee{\end{equation}}
\def\bea{\begin{eqnarray}}
\def\eea{\end{eqnarray}}

\def\HH{{\cal H}}

\def\ph{{\phi_{1}}}

\def\dotvarph{{\dot \varphi_{1}}}

\def\varph{{\varphi_{1}}}
\def\dotvarph{{\dot \varphi_{1}}}

\def\phsec{{\phi_{2}}}
\def\varphsec{{\varphi_{2}}}
\def\dotvarphsec{{\dot \varphi_{2}}}
\def\ps{{\psi_{1}}}
\def\dotps{{\dot \psi_{1}}}
\def\pssec{{\psi_{2}}}
\def\dotps{{\dot{\psi}_{1}}}
\def\dotpssec{{\dot{\psi}_{2}}}

\def\drho{{\delta \rho_{1}}}
\def\drhosec{{\delta \rho_{2}}}
\def\dotdrho{{\dot{\delta \rho}_{1}}}

\def\dP{{\delta P_{1}}}
\def\dPsec{{\delta P_{2}}}
\def\zet{{\zeta_{1}}}
\def\zetsec{{\zeta_{2}}}
\def\dotzet{{\dot{\zeta}_{1}}}
\def\dotzetsec{{\dot{\zeta}_{2}}}
\def\dPnad{{{\delta P}_{\rm 1, nad}}}
\def\dPnadsec{{{\delta P}_{\rm 2, nad}}}

\begin{document}



\title{On the conservation of second-order cosmological 
perturbations in a scalar field dominated universe}

\author{Filippo Vernizzi}
\affiliation{ Helsinki
Institute of Physics, P.O. Box 64, FIN-00014, University of
Helsinki, Finland}


\vspace{1cm}

\begin{abstract}
We discuss second-order cosmological per\-tur\-ba\-tions on
super-Hub\-ble scales, in a scalar field dominated universe, such
as during single field inflation. In this contest we show that the
gauge-invariant curvature perturbations defined on uniform
density and comoving hypersurfaces coincide and that perturbations
are adiabatic in the large scale limit. Since it has been recently
shown that the uniform density curvature perturbation is conserved on
large scales if perturbations are adiabatic, we conclude that both
the uniform density and comoving curvature perturbations at second-order,
in a scalar field dominated universe, are conserved. Finally, in the light 
of this result,
we comment on the variables recently used in the literature to compute 
non-Gaussianities.
\end{abstract}

\maketitle

\date{\today}


The study of second-order perturbation theory has recently become
important \cite{MW,A,Rigopoulus,Nakamura,Noh} (see e.g., also
\cite{Mukhanov,Bruni} for earlier work), especially because
primordial non-Gaussianities generated by inflation are typically
only of second-order level \cite{A,M,JP,Antti,Bartjan,review}. In the
study of cosmological perturbations and non-Gaussianities it is
very useful to establish results in terms of quantities that are
conserved on large scales, i.e., on super-Hubble scales. At linear
order, it is well known that the gauge invariant curvature
perturbations defined on uniform density and comoving
hypersurfaces -- the so called uniform density and comoving curvature
perturbations -- coincide and are both conserved on large scales
for adiabatic perturbations \cite{B}. On the other hand, when the
universe is dominated by a single scalar field, such as during
inflation, one can easily show that perturbations are adiabatic on
large scales \cite{KS,Bassett}. By adiabatic we intend that the
entropy perturbation (defined below) vanishes.

At second-order some results have been recently established on the
conservation properties of these variables. In \cite{MW} it has
been shown that it is possible to define a gauge invariant
quantity, the second-order generalization of the uniform density curvature
perturbation, which is conserved if the entropy perturbation of
the content of the universe vanishes. Furthermore, in \cite{M} two
conserved (although not gauge invariant) quantities (originally
defined in \cite{SB}) have been used to compute the level of
non-Gaussianities produced from inflation. Gauge-invariant
perturbation theory has also been studied in \cite{Noh}.

In this note we show, using the Einstein energy and momentum
constraint equations at perturbed second-order, that the gauge
invariant uniform density and comoving curvature perturbations $\zeta$ and
${\cal R}$, given as the sum of a first and a second-order
contribution to be defined below, 
\be \zeta = \zet + {\rm
\frac{1}{2}} \zetsec,  \quad \quad 
{\cal R} = {\cal R}_1 + {\rm
\frac{1}{2}} {\cal R}_2 , 
\ee
coincide (up to a sign)  on large
scales,
\be\zeta+{\cal R}
\approx 0.\ee Furthermore, we generalize at second-order the
well-known first-order result that the non-adiabatic entropy
perturbation of a scalar field on large scales is proportional to
the sum of the uniform density and comoving curvature perturbations (see
e.g., \cite{KS}), and thus vanishes,
\be
{\cal S} = \HH \left( \frac{\delta P}{\dot P} - \frac{\delta
\rho}{\dot \rho} \right)= -\frac{1-c_s^2}{c_s^2} (\zeta + {\cal
R}) \approx 0,
\ee
where $\rho$ and $P$ are the density and pressure of the field,
\be
\delta \rho= \drho +{\rm \frac{1}{2}} \drhosec, \quad \quad 
\delta
P= \dP +{\rm \frac{1}{2}} \dPsec,
\ee
are their perturbations up to
second-order, and 
\be 
c_s^2=\dot P/ \dot \rho \label{sos}
\ee
is the adiabatic speed of sound
of the field. Therefore, since the conservation of $\zeta$ at
second-order has been shown to hold under the condition that the
entropy perturbation of the content of the universe vanishes,
i.e., when ${\cal S}=0$ \cite{MW}, our conclusion implies that
both the uniform density and the comoving curvature perturbations are
conserved on super-Hubble scales when the universe is dominated by
a single minimally coupled scalar field,
\be
\zeta \approx -{\cal R}, \quad \dot \zeta \approx - \dot {\cal R}
\approx 0.
\ee

The advantage of using a conserved curvature perturbation to
compute non-Gaussianities produced in single field inflation, as it has been
done in \cite{M}, is
clear: since the comoving curvature perturbation is conserved on
super-Hubble scales, non-Gaussianities in this variable can only
be generated {\em inside the Hubble radius} due to the
self-coupling of the inflaton fluctuations in the vacuum, and are
later conserved in their super-Hubble evolution \cite{M}. Only the
presence of more than one field, which makes 
perturbations to be non-adiabatic, can source curvature
perturbations and generate non-Gaussianities on large scales.
Examples of this case have been discussed in \cite{Riotto} using
the uniform density curvature perturbation. 

We begin with a review of the first-order results. Then we define
the gauge invariant second-order uniform density and comoving curvature
perturbations for a scalar field and show that the source of their
evolution equation, the non-adiabatic pressure perturbation,
vanishes on large scales. By virtue of the results of \cite{MW},
we conclude that these variables are conserved. Finally, in the light 
of our results, we comment 
on the different variables used by Maldacena \cite{M} and Acquaviva et al.\ 
\cite{A} in computing non-Gaussianities from single field inflation.
The equality
symbol ``$=$'' means here {\em equal on all scales}, while
``$\approx$'' stems for {\em equal only on super-Hubble scales}.

We work with conformal time and use the perturbed
Friedmann-Lema\^{\i}tre-Robertson-Walker metric in the so called 
{\em generalized
longitudinal gauge} \cite{A},
\be
ds^2 = a^2 \left\{ -(1+2 \phi) d\eta^2
 + \left[ (1-2 \psi) \delta_{ij} + \chi_{ij}  \right] dx^i
dx^j \right\},
\ee
where we can expand the (spatially dependent) metric perturbation
variables at first and second-order, $\phi=\ph + {\rm
\frac{1}{2}}\phsec$ and $\psi=\ps +{\rm \frac{1}{2}}\pssec$, while
$\chi_{ij}={\rm \frac{1}{2}} \chi_{2,ij}$ is only second-order
and divergence-free, $\partial_i \chi^i_j=0$. A dot stems for a
derivative with respect to conformal time and $\HH=\dot a/a$. The
universe is dominated by a perturbed scalar field with background
value $\varphi=\varphi(\eta)$, energy density $\rho
=\dot \varphi^2/(2a^2)+a^2V$, and pressure $P
=\dot \varphi^2/(2a^2)-a^2V$. The scalar field
perturbation can be expanded into  first and  second-order
contributions, $\delta \varphi=\delta \varph + {\rm
\frac{1}{2}}\delta \varphsec$.

The Friedmann equation is 
\be3\HH^2 = \kappa^2 \left({\rm \frac{1}{2}}\dot
\varphi^2+a^2V \right) , 
\ee
where $\kappa^2=8 \pi G$. We shall use the
background scalar field evolution equation, $\dot \rho =-3 \HH
\dot \varphi^2/a^2$, or 
\be
\ddot \varphi + 2\HH \dot \varphi +V'=0,
\ee
where $V'$ is the derivative of the scalar field potential with
respect to the field. We shall also repeatedly make use of the
relation $\kappa^2 \dot \varphi^2 /2 = \HH^2 -\dot \HH$, which can
be derived from the Friedmann equation. The adiabatic speed of
sound of the scalar field defined in Eq.~(\ref{sos}) 
is 
\be
c_s^2 = 1-2V'\dot \varphi/\dot
\rho. \label{pippetto}
\ee
\\

{\em First-order perturbations.}
At first-order, the perturbed energy constraint reads
(see e.g., \cite{KS})
\be
6 \HH (\dotps+ \HH \ph) -2\Delta \ps= -\kappa^2 \drho a^2,
\label{en1}
\ee
with 
\be
\drho =  \frac{1}{a^2} (\dot \varphi \delta \dotvarph -\ph \dot
\varphi^2) + V'\delta \varph , \label{drho}
\ee
$\Delta=\partial^i \partial_i$, and the momentum constraint is 
\be
2(\dotps + \HH \ph) = \kappa^2 \dot \varphi \delta \varph.
\label{mom1}
\ee

The uniform density curvature perturbation and the comoving
curvature perturbation are defined as \cite{B,WMLL}
\be \zet=-\ps -\HH
\frac{\drho}{\dot \rho}, \quad \quad {\cal R}_1=\ps +\HH
\frac{\delta \varph}{\dot \varphi}. \label{def}
\ee
By definition,
both these variables reduce to the curvature perturbation $\psi_1$ (up to
a sign) on setting $\delta \rho_1=0$ and $\delta \varphi_1= 0$, respectively.
Indeed, for a scalar field, 
the comoving hypersurfaces are the hypersurfaces of uniform field.

On using the energy and momentum constraints, Eqs.~(\ref{en1}) and
(\ref{mom1}), at first-order one can show that uniform density
and uniform field (or comoving) hypersurfaces coincide on large
scales. Indeed we have,
\be
\frac{\drho}{\dot \rho} - \frac{\delta \varph}{\dot \varphi} =
-\frac{\Delta \ps}{3\HH(\HH^2-\dot \HH)} \approx 0. \label{hyp}
\ee
Thus,  uniform density and comoving curvature perturbations coincide on
large scales, up to a sign,
\be
\zet \approx-{\cal R}_1. \label{related1}
\ee
Equation (\ref{hyp}), together with (\ref{drho}), also yields
\be
\delta \dotvarph  \approx \ph \dot \varphi + (\ddot \varphi / \dot
\varphi) \delta \varph -\HH \delta \varph. \label{using}
\ee

{\em For a general fluid} and in absence of anisotropic stress,
both curvature perturbation variables are sourced by the
non-adiabatic pressure perturbation \footnote{The non-adiabatic
perturbation is proportional to the entropy perturbation, $ \delta
P_{\rm nad} = \dot P {\cal S}$.},
\be
\dPnad=\dP -c_s^2 \drho. \label{dPdef}
\ee
Indeed, their evolution is given by \cite{WMLL}
\bea
\dotzet&=&\frac{3\HH^2}{\dot \rho} \dPnad +\frac{1}{3}\Delta
v_1
\approx \frac{3\HH^2}{\dot \rho} \dPnad,\\
\dot {\cal R}_1 &=& -\frac{3\HH^2}{\dot \rho} \dPnad + 3 \HH
c_s^2 (\zet + {\cal R}_1)  
\nonumber \\
&\approx&-\frac{3\HH^2}{\dot \rho}
\dPnad,
\eea
where $v_1$ is the scalar component of the three-velocity of
the fluid (for a scalar field $v_1=\delta \varph / \dot \varphi$). 
However, {\em for a scalar field} 
\be
\dP = \frac{1}{a^2} (\varphi
\delta \varph -\ps \dot \varphi^2) - V' \delta \varph =\drho 
-2V' \delta \varph , 
\ee
and the definition (\ref{dPdef}) with Eq.~(\ref{pippetto}) yields
\cite{KS}
\be
\HH \frac{\dPnad}{\dot \rho} = -(1-c_s^2) (\zet + {\cal R}_1)
\approx 0, \label{use}
\ee
where we have used Eq.~(\ref{related1}) for the second equality.
This implies that both uniform density and comoving curvature
perturbations are conserved on large scales
\be
\dotzet \approx - \dot {\cal R}_1 \approx 0. \label{cons1}
\ee
\\

{\em Second-order perturbations.}
Now we generalize these results to second-order. For simplicity we
use that $\ps = \ph$ in a scalar field dominated universe, which
yields from the traceless $ij$ part of the Einstein equations, 
and from the fact
that the scalar field anisotropic stress vanishes at first-order. It
is not a necessary condition for our results but it considerably
simplifies their proof. Note, however, that at second-order, 
in the generalized 
longitudinal gauge, we have $\pssec \neq\phsec$
\cite{A}, which makes second-order calculations quite involved.

We shall consider the perturbed energy and momentum
constraints only on large scales. At second-order  they read
\cite{A}
\be
6 \HH (\dotpssec + \HH \phsec) -24 \HH^2 \ps^2 -6 \dotps{}^2
\approx -\kappa^2 \delta \rho_2 a^2, \label{en2}
\ee
and
\bea
\partial^i ( \dotpssec+ \HH \phsec) + 2 \partial^i (\ps \dotps)
+ 6 \ps \partial^i \dotps \ \ \ \ \nonumber \\  \approx \kappa^2 \left[
\frac{1}{2} \dot \varphi
\partial^i \delta \varphsec +(\delta \dotvarph 
+ 2 \dot \varphi \ps )\partial^i \delta \varph \right]. \label{mom2}
\eea
[The large scale definition of $\delta \rho_2$ is given in Eq.~(\ref{drho2}) below.]

We can use the first-order momentum constraint to write the last
term in the first line of Eq.~(\ref{mom2}) as
\be
6 \ps \partial^i \dotps \approx 3 \kappa^2 \ps \dot \varphi
\partial^i \delta \varph -3 \HH \partial^i (\ps{}^2),
\ee
and we can replace $\delta \dotvarph$ in the second line of
Eq.~(\ref{mom2})  by the expression (\ref{using}). By this
replacement we find a much simpler form for the perturbed momentum
constraint at second-order, that holds only on large scales, and does 
not involve spatial derivatives,
\bea
\dotpssec + \HH \phsec + 2 \ps \dotps-3 \HH \ps{}^2 \quad \quad \quad     
  \nonumber \\
\approx
\frac{\kappa^2}{2} \left[ \dot \varphi \delta \varphsec + \left(
\frac{\ddot \varphi}{\dot \varphi} -\HH  \right)\delta \varph^2
\right].
\eea
We can combine this equation and the second-order energy
constraint, Eq.~(\ref{en2}), to eliminate the term $\dot \pssec +
\HH \phsec$ obtaining
\bea
\frac{\delta \rho_2}{\dot \rho} - \frac{\delta \varphsec}{\dot
\varphi}  \approx  \left( \frac{\ddot \rho}{\dot \rho}   -
\frac{\ddot \varphi}{\dot \varphi}\right) \frac{\delta
\varph^2}{\dot \varphi^2}, \label{diff}
\eea
where we have made use of Eq.~(\ref{mom1}). This relation between
the energy density and the field perturbations is the second-order
analog of Eq.~(\ref{hyp}) and we shall use it below to show the
equivalence between the two curvature perturbations.

The second-order uniform density curvature perturbation has been defined
in \cite{MW}. On large scales it reads
\bea
\zetsec \approx - \pssec - \HH\frac{\drhosec}{\dot \rho} + 2 \HH
\frac{\dotdrho \drho }{\dot \rho^2}  
+ 2
\frac{\drho}{\dot \rho} (\dotps+2\HH \ps) \nonumber \\  + \frac{\drho^2}{\dot
\rho^2} \left(  \dot \HH +2 \HH^2- \HH \frac{\ddot \rho}{\dot
\rho} \right).\quad \quad  \label{zetasecdef}
\eea
We define the second-order comoving curvature perturbation
as 
\bea
{\cal R}_2 \approx \pssec +\HH\frac{\delta \varphsec}{\dot
\varphi} - 2 \HH \frac{\delta \dotvarph \delta \varph }{\dot
\varphi^2} - 2 \frac{\delta \varph}{\dot \varphi} (\dotps +2 \HH \ps) 
\nonumber
\\ -
\frac{\delta \varph^2}{\dot \varphi^2} \left(  \dot \HH + 2\HH^2 -
\HH \frac{\ddot \varphi}{\dot \varphi} \right). \quad \quad  \label{Rsecdef}
\eea
By setting $\delta \varphi=0$ one can check that the comoving curvature 
perturbation here defined 
simply reduces to the curvature $\psi$, up to second-order, 
${\cal R} \simeq \psi$.
The variables $\zeta_2$ and ${\cal R}_2$ 
are defined up to small scale terms. One can
check that they are gauge invariant on large scales.

The sum of these two variables becomes easy to compute once
Eq.~(\ref{diff}) is known. Indeed, from Eq.~(\ref{diff}),
\bea \zetsec + {\cal R}_2
= - 2(\dotzet+2\HH \zet) \frac{\drho}{\dot \rho} -2 ( \dot {\cal
R}_1 +2 \HH {\cal R}_1) \frac{\delta \varph}{\dot \varphi}
\nonumber \\+ \left(\frac{\dot \HH}{\HH}+2 \HH - \frac{\ddot
\rho}{\dot \rho} \right) \left(\zet + {\cal R}_1 \right)
\left(\frac{\delta \varph}{\dot
\varphi} + \frac{\drho}{\dot \rho}\right) \approx 0, \nonumber \\
\label{related}
\eea
where we have used Eqs.~(\ref{related1}) and (\ref{cons1}) for the last
equality. Thus
$\zetsec \approx -{\cal R}_2$.

The large scale evolution equation of the second-order uniform
density curvature perturbation has been derived in \cite{MW}. On large
scales, {\em for a general fluid}, it yields
\bea 
\dotzetsec \approx \frac{3\HH^2}{\dot
\rho} \left(\dPnadsec - 2 \frac{\drho}{ \dot \rho} \dot{\delta P}_{1,\rm nad}
\right) \nonumber \\
+ \left(\frac{6 \HH}{\dot \rho} \dPnad + 4\zet \right)
\dot \zet, \label{cons}
\eea
where the non-adiabatic second-order pressure perturbation is
defined as \cite{MW}
\be
\dPnadsec = \delta P^{(2)} -c_s^2 \delta \rho^{(2)} - \dot c_s^2
\frac{\drho^2}{\dot \rho}.
\ee
Equation (\ref{related}) implies
\bea
\dot {\cal R}_2 \approx -\frac{3\HH^2}{\dot \rho}
\left(\dPnadsec - 2 \frac{\drho}{ \dot \rho} \dot{\delta P}_{1,\rm nad} \right)
\nonumber \\
+\left( \frac{6 \HH}{\dot \rho} \dPnad - 4{\cal R}_1 \right)
\dot {\cal R}_1 . \label{cons2}
\eea
{\em For a scalar field} $\dPnad \approx 0$, from Eq.~(\ref{use}).
Thus $\dPnadsec$ is the only possible non-vanishing contribution
on the right hand side of Eqs.~(\ref{cons}) and (\ref{cons2}). Now
we show that ${\dPnadsec}$ vanishes as well.

From the large scale definition of $\drhosec$ and $\delta P_2$
\cite{A},
\bea
\drhosec  \approx \frac{1}{a^2} [\dot \varphi \delta \dotvarphsec - \phsec
\dot \varphi^2   + \delta
\dotvarph{}^2 + 4 \ps \dot \varphi (  \ps \dot \varphi - \delta
\dotvarph)]  \nonumber \\ + V' \delta \varphsec + V'' \delta \varph^2, 
\quad \quad \quad \label{drho2}\\
\delta P_2  \approx  \frac{1}{a^2} [\dot \varphi \delta \dotvarphsec -
\phsec \dot \varphi^2  +
\delta \dotvarph{}^2 + 4 \ps \dot \varphi (  \ps \dot \varphi -
\delta \dotvarph)] \nonumber 
\\  -  V' \delta \varphsec - V'' \delta \varph^2, \quad \quad \quad 
\eea
we find $\delta P_2 \approx \drhosec -2 V'a^2 \delta \varphsec
-2 V'' a^2 \delta \varph^2$. 
This yields
\bea
\HH \frac{\dPnadsec}{\dot \rho}&=& (1-c_s^2) \HH \left(
\frac{\drhosec}{\dot \rho} - \frac{\delta \varphsec}{\dot \varphi}
\right) - \dot c_s^2 \HH \frac{\drho^2}{\dot \rho^2} \nonumber \\
&&+ \HH \left[(1-c_s^2) \left(\frac{\ddot \varphi}{\dot \varphi }
- \frac{\ddot \rho}{\dot \rho} \right) + \dot c_s^2 \right] \frac{\delta \varph^2}{\dot \varphi^2}\nonumber \\
&\approx& - (1-c_s^2) (\zetsec + {\cal R}_2) \approx
0,\label{dPnad2}
\eea
by virtue of Eq.~(\ref{related}). The second-order non-adiabatic
pressure perturbation of a dominating scalar field vanishes on
large scales, ${\dPnadsec} \approx 0$, i.e., single scalar field
perturbations are adiabatic on large scales to second-order.
Equations (\ref{related}) and (\ref{dPnad2}) are our main results.
As a consequence, $\zeta_2$ and ${\cal R}_2$ as defined by
Eqs.~(\ref{zetasecdef}) and (\ref{Rsecdef}) are both conserved,
\be
 \dotzetsec \approx - \dot {\cal R}_2 \approx 0.
\ee

A final remark concerning the literature on non-Gaussianities
is in order here. The second-order comoving curvature perturbation
definition of Eq.~(\ref{Rsecdef}) 
differs from the one given by Acquaviva et al.\ 
in Ref.~\cite{A} and used in \cite{Rigopoulus,Antti}.
Our variable is related to their variable by
\be
{\cal R}_{\rm 2, A} =  {\cal R}_{2}
+ \frac{(\dot {\cal R}_{1} + 2\HH {\cal R}_{1})^2}{(\dot{\HH} +
2 \HH^2 -\HH \ddot \varphi / \dot \varphi)} .
\ee
Indeed ${\cal R}_{\rm 2, A}$ is not the curvature perturbation
defined on comoving hypersurfaces, 
as one can verify by setting 
$\delta \varphi=0$ in their definition, and 
fails to be conserved on large
scales. 
For adiabatic perturbations we can easily derive its evolution. 
By using that ${\cal R}_2$ is conserved on large scales we find
\be
\dot {\cal R}_{\rm 2, A} \approx  \frac{4 (2 \dot \epsilon
-\dot \eta)}{(2-2 \epsilon +\eta)^2} {\cal R}_{1}^2 \sim (2
\dot \epsilon -\dot \eta) {\cal R}_{1}^2, \label{final}
\ee
where 
$\epsilon=(M^2_{\rm Pl}/2)(V'/V)^2$ and $\eta=M^2_{\rm
Pl}V''/V$ are the standard slow-roll parameters, and we have 
expanded to {\em second} order in the 
slow-roll approximation {\em only} in the second equality.
Equation (\ref{final}) yields the solution
\be
{\cal R}_{\rm 2, A} = (2 \epsilon -\eta){\cal R}^2_{1}+ {\cal C}, 
\ee
where $\cal C$ is some integration constant which has to be
specified by the initial 
conditions, i.e., by some {\em sub-Hubble} scale physics. 
Hence, our result appears to disagree
with Eq.~(63) of \cite{A}.

Furthermore, the variable ${\cal R}_{\rm 2, A}$ has been used in \cite{A} to 
compute the non-Gaussianities from single field inflation. 
However, since ${\cal R}_{\rm 2, A}$ is not conserved on large scales, the
non-Gaussianities  
obtained in \cite{A} should disagree
with those found by Maldacena \cite{M}, who used a conserved quantity.
Indeed, one
can check
that the second-order variables used by Maldacena in \cite{M}
are related to our variable by
$\zeta^{(\delta \varphi)}_{\rm M} = -{\cal R}-
{\cal R}^2$ in the flat gauge, where $\psi =0$, and  by
$\zeta^{(\psi)}_{\rm M} = {\cal R}+
{\cal R}^2$
in the uniform field (or comoving) gauge, where $\delta \varphi=0$.
These variables are thus perfectly conserved for adiabatic perturbations.

\vskip0.2cm {\em Note added:} While writing this note, a proof of
the conservation of the uniform density and comoving
curvature perturbations on large scales without perturbative
expansion has been given in
\cite{K}. A geometric and  
fully non-linear generalization of the uniform density and comoving
curvature perturbations
will be presented in
\cite{DF}.
\\

\vskip0.2cm {\bf Acknowledgments:} I acknowledge useful
discussions with Gianluca Calcagni, Karim Malik, and Antti V\"aihk\"onen.


\end{document}